\def\unredoffs{} \def\redoffs{\voffset=-.40truein\hoffset=-.40truein}
\def\speclscape{}

\newbox\leftpage \newdimen\fullhsize \newdimen\hstitle \newdimen\hsbody
\tolerance=1000\hfuzz=2pt
\catcode`\@=11 % This allows us to modify PLAIN macros.
\def\bigans{b }
%%%% \message{ big or little (b/l)? }\read-1 to\answ
\def\answ{b }
\ifx\answ\bigans\message{(This will come out unreduced.}
%\magnification=900
\unredoffs\baselineskip=16pt plus 2pt minus 1pt
\hsbody=\hsize \hstitle=\hsize %take default values for unreduced format
\else\message{(This will be reduced.} \let\l@r=L
%\magnification=1000\baselineskip=16pt plus 2pt minus 1pt \vsize=7truein
\redoffs \hstitle=8truein\hsbody=4.75truein\fullhsize=10truein\hsize=\hsbody
\output={\ifnum\pageno=0 %%% This is the HUTP version
   \shipout\vbox{\speclscape{\hsize\fullhsize\makeheadline}
     \hbox to \fullhsize{\hfill\pagebody\hfill}}\advancepageno
   \else
   \almostshipout{\leftline{\vbox{\pagebody\makefootline}}}\advancepageno
   \fi}
\def\almostshipout#1{\if L\l@r \count1=1 \message{[\the\count0.\the\count1]}
       \global\setbox\leftpage=#1 \global\let\l@r=R
  \else \count1=2
   \shipout\vbox{\speclscape{\hsize\fullhsize\makeheadline}
       \hbox to\fullhsize{\box\leftpage\hfil#1}}  \global\let\l@r=L\fi}
\fi
%---------------------------------------------------------------------
%
\newcount\yearltd\yearltd=\year\advance\yearltd by -1900

%
%  restores pagenumbers
%
%       use following instead of \Date on the preliminary draft,
%       puts date/time on each page in big mode, writes labels in margins

\def\draftmode{\message{ DRAFTMODE }\def\draftdate{{\rm preliminary draft:
\number\month/\number\day/\number\yearltd\ \ \hourmin}}%
\headline={\hfil\draftdate}\writelabels\baselineskip=20pt plus 2pt minus 2pt
  {\count255=\time\divide\count255 by 60 \xdef\hourmin{\number\count255}
   \multiply\count255 by-60\advance\count255 by\time
   \xdef\hourmin{\hourmin:\ifnum\count255<10 0\fi\the\count255}}}
%       use \nolabels to get rid of eqn, ref, and fig labels in draft mode
\def\nolabels{\def\wrlabeL##1{}\def\eqlabeL##1{}\def\reflabeL##1{}}
\def\writelabels{\def\wrlabeL##1{\leavevmode\vadjust{\rlap{\smash%
{\line{{\escapechar=` \hfill\rlap{\sevenrm\hskip.03in\string##1}}}}}}}%
\def\eqlabeL##1{{\escapechar-1\rlap{\sevenrm\hskip.05in\string##1}}}%
\def\reflabeL##1{\noexpand\llap{\noexpand\sevenrm\string\string\string##1}}}
\nolabels
%
% tagged sec numbers
\global\newcount\secno \global\secno=0
\global\newcount\meqno \global\meqno=1
\def\newsec#1{\global\advance\secno by1\message{(\the\secno. #1)}
%\ifx\answ\bigans \vfill\eject \else \bigbreak\bigskip \fi  %if desired
\global\subsecno=0\eqnres@t\noindent{\boldLARGE \the\secno. #1} \writetoca{{\secsym}
{#1}}\par\nobreak\medskip\nobreak}
\def\eqnres@t{\xdef\secsym{\the\secno.}\global\meqno=1\bigbreak\bigskip}
\def\sequentialequations{\def\eqnres@t{\bigbreak}}\xdef\secsym{}
\global\newcount\subsecno \global\subsecno=0
\def\subsec#1{\global\advance\subsecno by1\message{(\secsym\the\subsecno. #1)}
\ifnum\lastpenalty>9000\else\bigbreak\fi \noindent{\boldlarge \secsym\the\subsecno. #1}\writetoca{\string\quad
{\secsym\the\subsecno.} {#1}}\par\nobreak\medskip\nobreak}
\def\appendix#1#2{\global\meqno=1\global\subsecno=0\xdef\secsym{\hbox{#1.}}
\bigbreak\bigskip\noindent{\bf Appendix #1. #2}\message{(#1. #2)}
\writetoca{Appendix {#1.} {#2}}\par\nobreak\medskip\nobreak}
%
%       \eqn\label{a+b=c}   gives displayed equation, numbered
%               consecutively within sections.
%     \eqnn and \eqna define labels in advance (of eqalign?)
%
\def\eqnn#1{\xdef #1{(\secsym\the\meqno)}\writedef{#1\leftbracket#1}%
\global\advance\meqno by1\wrlabeL#1}
\def\eqna#1{\xdef #1##1{\hbox{$(\secsym\the\meqno##1)$}}
\writedef{#1\numbersign1\leftbracket#1{\numbersign1}}%
\global\advance\meqno by1\wrlabeL{#1$\{\}$}}
\def\eqn#1#2{\xdef #1{(\secsym\the\meqno)}\writedef{#1\leftbracket#1}%
\global\advance\meqno by1$$#2\eqno#1\eqlabeL#1$$}
%
%            footnotes
\newskip\footskip\footskip14pt plus 1pt minus 1pt %sets footnote baselineskip
\def\footnotefont{\ninepoint}\def\f@t#1{\footnotefont #1\@foot}
\def\f@@t{\baselineskip\footskip\bgroup\footnotefont\aftergroup\@foot\let\next}
\setbox\strutbox=\hbox{\vrule height9.5pt depth4.5pt width0pt}
\global\newcount\ftno \global\ftno=0
\def\foot{\global\advance\ftno by1\footnote{$^{\the\ftno}$}}
%
%say \footend to put footnotes at end
%will cause problems if \ref used inside \foot, instead use \nref before
\newwrite\ftfile
\def\footend{\def\foot{\global\advance\ftno by1\chardef\wfile=\ftfile
$^{\the\ftno}$\ifnum\ftno=1\immediate\openout\ftfile=foots.tmp\fi%
\immediate\write\ftfile{\noexpand\smallskip%
\noexpand\item{f\the\ftno:\ }\pctsign}\findarg}%
\def\footatend{\vfill\eject\immediate\closeout\ftfile{\parindent=20pt
\centerline{\bf Footnotes}\nobreak\bigskip\input foots.tmp }}}
\def\footatend{}
%
%     \ref\label{text}
% generates a number, assigns it to \label, generates an entry.
% To list the refs on a separate page,  \listrefs
%
\global\newcount\refno \global\refno=1
\newwrite\rfile
\def\ref{[\the\refno]\nref}
\def\nref#1{\xdef#1{[\the\refno]}\writedef{#1\leftbracket#1}%
\ifnum\refno=1\immediate\openout\rfile=refs.tmp\fi
\global\advance\refno by1\chardef\wfile=\rfile\immediate
\write\rfile{\noexpand\item{#1\ }\reflabeL{#1\hskip.31in}\pctsign}\findarg}
%   horrible hack to sidestep tex \write limitation
\def\findarg#1#{\begingroup\obeylines\newlinechar=`\^^M\pass@rg}
{\obeylines\gdef\pass@rg#1{\writ@line\relax #1^^M\hbox{}^^M}%
\gdef\writ@line#1^^M{\expandafter\toks0\expandafter{\striprel@x #1}%
\edef\next{\the\toks0}\ifx\next\em@rk\let\next=\endgroup\else\ifx\next\empty%
\else\immediate\write\wfile{\the\toks0}\fi\let\next=\writ@line\fi\next\relax}}
\def\striprel@x#1{} \def\em@rk{\hbox{}}
\def\lref{\begingroup\obeylines\lr@f}
\def\lr@f#1#2{\gdef#1{\ref#1{#2}}\endgroup\unskip}
\def\semi{;\hfil\break}
\def\addref#1{\immediate\write\rfile{\noexpand\item{}#1}} %now unnecessary
\def\startrefs#1{\immediate\openout\rfile=refs.tmp\refno=#1}
\def\xref{\expandafter\xr@f}\def\xr@f[#1]{#1}
\def\refs#1{\count255=1[\r@fs #1{\hbox{}}]}
\def\r@fs#1{\ifx\und@fined#1\message{reflabel \string#1 is undefined.}%
\nref#1{need to supply reference \string#1.}\fi%
\vphantom{\hphantom{#1}}\edef\next{#1}\ifx\next\em@rk\def\next{}%
\else\ifx\next#1\ifodd\count255\relax\xref#1\count255=0\fi%
\else#1\count255=1\fi\let\next=\r@fs\fi\next}
%

%
% this is ugly, but moore insists
\newwrite\ffile\global\newcount\figno \global\figno=1
\def\fig{fig.~\the\figno\nfig}
\def\nfig#1{\xdef#1{fig.~\the\figno}%
\writedef{#1\leftbracket fig.\noexpand~\the\figno}%
\ifnum\figno=1\immediate\openout\ffile=figs.tmp\fi\chardef\wfile=\ffile%
\immediate\write\ffile{\noexpand\medskip\noexpand\item{Fig.\ \the\figno. }
\reflabeL{#1\hskip.55in}\pctsign}\global\advance\figno by1\findarg}
\def\xfig{\expandafter\xf@g}\def\xf@g fig.\penalty\@M\ {}
\def\figs#1{figs.~\f@gs #1{\hbox{}}}
\def\f@gs#1{\edef\next{#1}\ifx\next\em@rk\def\next{}\else
\ifx\next#1\xfig #1\else#1\fi\let\next=\f@gs\fi\next}
\newwrite\lfile
{\escapechar-1\xdef\pctsign{\string\%}\xdef\leftbracket{\string\{}
\xdef\rightbracket{\string\}}\xdef\numbersign{\string\#}}

\def\writestop{\def\writestoppt{\immediate\write\lfile{\string\pageno%
\the\pageno\string\startrefs\leftbracket\the\refno\rightbracket%
\string\def\string\secsym\leftbracket\secsym\rightbracket%
\string\secno\the\secno\string\meqno\the\meqno}\immediate\closeout\lfile}}
\def\writestoppt{}\def\writedef#1{}
\def\seclab#1{\xdef #1{\the\secno}\writedef{#1\leftbracket#1}\wrlabeL{#1=#1}}
\def\subseclab#1{\xdef #1{\secsym\the\subsecno}%
\writedef{#1\leftbracket#1}\wrlabeL{#1=#1}}
\newwrite\tfile \def\writetoca#1{}
\def\leaderfill{\leaders\hbox to 1em{\hss.\hss}\hfill}
%   use this to write file with table of contents
\def\writetoc{\immediate\openout\tfile=toc.tmp
    \def\writetoca##1{{\edef\next{\write\tfile{\noindent ##1
    \string\leaderfill {\noexpand\number\pageno} \par}}\next}}}
%       and this lists table of contents on second pass

%
\catcode`\@=12 % at signs are no longer letters
%
%   Unpleasantness in calling in abstract and title fonts
\edef\tfontsize{\ifx\answ\bigans scaled\magstep3\else scaled\magstep4\fi}
 \tfontsize  \tfontsize
 \tfontsize \font\titlei=cmmi10 \tfontsize
\font\titleis=cmmi7 \tfontsize \font\titleiss=cmmi5 \tfontsize
\font\titlesy=cmsy10 \tfontsize \font\titlesys=cmsy7 \tfontsize
\font\titlesyss=cmsy5 \tfontsize  \tfontsize
\skewchar\titlei='177 \skewchar\titleis='177 \skewchar\titleiss='177
\skewchar\titlesy='60 \skewchar\titlesys='60 \skewchar\titlesyss='60
 \ifx\answ\bigans\else scaled\magstep1\fi
\ifx\answ\bigans\else

 \font\absi=cmmi10 scaled\magstep1
\font\absis=cmmi7 scaled\magstep1 \font\absiss=cmmi5 scaled\magstep1
\font\abssy=cmsy10 scaled\magstep1 \font\abssys=cmsy7 scaled\magstep1
\font\abssyss=cmsy5 scaled\magstep1 
\skewchar\absi='177 \skewchar\absis='177 \skewchar\absiss='177
\skewchar\abssy='60 \skewchar\abssys='60 \skewchar\abssyss='60
\fi
\font\ninerm=cmr9 \font\sixrm=cmr6 \font\ninei=cmmi9 \font\sixi=cmmi6
\font\ninesy=cmsy9 \font\sixsy=cmsy6 \font\ninebf=cmbx9
\font\nineit=cmti9 \font\ninesl=cmsl9 \skewchar\ninei='177
\skewchar\sixi='177 \skewchar\ninesy='60 \skewchar\sixsy='60
\def\ninepoint{\def\rm{\fam0\ninerm}% switch to footnote font
\textfont0=\ninerm \scriptfont0=\sixrm \scriptscriptfont0=\fiverm
\textfont1=\ninei \scriptfont1=\sixi \scriptscriptfont1=\fivei
\textfont2=\ninesy \scriptfont2=\sixsy \scriptscriptfont2=\fivesy
\textfont\itfam=\ninei \def\it{\fam\itfam\nineit}\def\sl{\fam\slfam\ninesl}%
\textfont\bffam=\ninebf \def\bf{\fam\bffam\ninebf}\rm}
%
%---------------------------------------------------------------------
%

\hyphenation{anom-aly anom-alies coun-ter-term coun-ter-terms}
\def\inv{^{\raise.15ex\hbox{${\scriptscriptstyle -}$}\kern-.05em 1}}

\def\Dsl{\,\raise.15ex\hbox{/}\mkern-13.5mu D} %this one can be subscripted
\def\dsl{\raise.15ex\hbox{/}\kern-.57em\partial}

 \def\Tr{{\rm Tr}}

 %pound sterling
\def\lspace{\ifx\answ\bigans{}\else\qquad\fi}
\def\lbspace{\ifx\answ\bigans{}\else\hskip-.2in\fi} % $$\lbspace...$$
\def\boxeqn#1{\vcenter{\vbox{\hrule\hbox{\vrule\kern3pt\vbox{\kern3pt
     \hbox{${\displaystyle #1}$}\kern3pt}\kern3pt\vrule}\hrule}}}
\def\mbox#1#2{\vcenter{\hrule \hbox{\vrule height#2in
         \kern#1in \vrule} \hrule}}  %e.g. \mbox{.1}{.1}
%   matters of taste
%\def\tilde{\widetilde} \def\bar{\overline} \def\hat{\widehat}
%
% some sample definitions
 \def\CO{{\cal O}} %     curly letters
\def\CA{{\cal A}}  \def\CF{{\cal F}} 
 \def\CH{{\cal H}}  
  \def\CD{{\cal D}} 
\def\e#1{{\rm e}^{^{\textstyle#1}}}

\def\darr#1{\raise1.5ex\hbox{$\leftrightarrow$}\mkern-16.5mu #1}
 %pound sterling

\def\half{{\textstyle{1\over2}}} %puts a small half in a displayed eqn
\def\roughly#1{\raise.3ex\hbox{$#1$\kern-.75em\lower1ex\hbox{$\sim$}}}

\def\np#1#2#3{Nucl. Phys. {\bf B#1} (#2) #3}
\def\pl#1#2#3{Phys. Lett. {\bf #1B} (#2) #3}

\def\anp#1#2#3{Ann. Phys. {\bf #1} (#2) #3}

\def\prep#1#2#3{Phys. Rep. {\bf #1} (#2) #3}

\def\cmp#1#2#3{Comm. Math. Phys. {\bf #1} (#2) #3}

\def\jhep#1#2#3{JHEP {\bf#1}(#2) #3}

\def\atmp#1#2#3{Adv.~Theor.~Math.~Phys.{\bf #1} (#2) #3}

%%%%%%%%%%%%%%%  Rublenye bukvy   %%%%%%%%%%%%%%%%%
\def\IB{\relax\hbox{$\inbar\kern-.3em{\rm B}$}}
\def\IC{\relax\hbox{$\inbar\kern-.3em{\rm C}$}}
\def\ID{\relax\hbox{$\inbar\kern-.3em{\rm D}$}}
\def\IE{\relax\hbox{$\inbar\kern-.3em{\rm E}$}}
\def\IF{\relax\hbox{$\inbar\kern-.3em{\rm F}$}}
\def\IG{\relax\hbox{$\inbar\kern-.3em{\rm G}$}}
\def\IGa{\relax\hbox{${\rm I}\kern-.18em\Gamma$}}
\def\IH{\relax{\rm I\kern-.18em H}}
\def\IK{\relax{\rm I\kern-.18em K}}
\def\IL{\relax{\rm I\kern-.18em L}}
\def\IP{\relax{\rm I\kern-.18em P}}
\def\IR{\relax{\rm I\kern-.18em R}}
\def\IZ{\relax\ifmmode\mathchoice{
\hbox{\cmss Z\kern-.4em Z}}{\hbox{\cmss Z\kern-.4em Z}}
{\lower.9pt\hbox{\cmsss Z\kern-.4em Z}}
{\lower1.2pt\hbox{\cmsss Z\kern-.4em Z}}
\else{\cmss Z\kern-.4em Z}\fi}
\def\II{\relax{\rm I\kern-.18em I}}

\def\ndt{{\noindent}}

%%%%%%%% Calligraphic letters  %%%%%%%%%%%%%

\def\CA{{\cal A}}

\def\CD{{\cal D}}
\def\CE{{\cal E}}
\def\CF{{\cal F}}

\def\CH{{\cal H}}

\def\CM{{\cal M}}
\def\CN{{\cal N}}
\def\CO{{\cal O}}

\def\CS{{\cal S}}

\def\CZ{{\cal Z}}

%%%%%%%%%%%% Derivatives  %%%%%%%%%%%
\def\p{\partial}
\def\pb{\bar{\partial}}

%%%%%%%%%%% letters with bar %%%%%%%%

%%%%%%%%%% Math symbols %%%%%%%%%%%%%

\def\Tr{{\rm Tr}}

%%%%%%%%%%%%%% Lie algebras %%%%%%%%%%%%%%%%%%%%%%

\def\inbar{\,\vrule height1.5ex width.4pt depth0pt}

\font\cmss=cmss10 \font\cmsss=cmss10 at 7pt

%%%%%%%%%%%% Greek %%%%%%%%%%%%
\def\a{{\alpha}}

\def\b{{\beta}}

\def\e{{\epsilon}}
\def\z{{\zeta}}
\def\ve{{\varepsilon}}

\def\m{{\mu}}
\def\n{{\nu}}
\def\u{{\Upsilon}}
\def\l{{\lambda}}
\def\s{{\sigma}}
\def\t{{\theta}}

\def\o{{\omega}}

%%%%%

\def\IF{{\bf F}}
\def\boxit#1{\vbox{\hrule\hbox{\vrule\kern8pt
\vbox{\hbox{\kern8pt}\hbox{\vbox{#1}}\hbox{\kern8pt}}
\kern8pt\vrule}\hrule}}
\def\mathboxit#1{\vbox{\hrule\hbox{\vrule\kern8pt\vbox{\kern8pt
\hbox{$\displaystyle #1$}\kern8pt}\kern8pt\vrule}\hrule}}
%%%%%%%Russian fonts%%%%%%%5

\chardef\tempcat=\the\catcode`\@ \catcode`\@=11
\def\cyracc{\def\u##1{\if \i##1\accent"24 i%
    \else \accent"24 ##1\fi }}
\newfam\cyrfam

%%%%% REFS %%%%%%%%%%

\def\lref{\begingroup\obeylines\lr@f}
\def\lr@f#1#2{\gdef#1{\ref#1{#2.}}\endgroup\unskip}

\lref\ikkt{N.~Ishibashi, H.~Kawai, Y.~Kitazawa, and A.~Tsuchiya,
\np{498}{1997}{467}, hep-th/9612115}

\lref\cds{A.~Connes, M.~Douglas, A.~Schwarz,
\jhep{9802}{1998}{003}} \lref\douglashull{M.~Douglas, C.~Hull,
``D-Branes and the noncommutative torus'', \jhep{9802}{1998}{008},
hep-th/9711165} \lref\wtnc{E.~Witten, \np{268}{1986}{253}}

\lref\gopakumarvafa{R.~Gopakumar, C.Vafa, hep-th/9809187,
hep-th/9812127}

\lref\wittenone{E.~Witten, hep-th/9403195}

\lref\cg{E.~Corrigan, P.~Goddard, ``Construction of instanton and
monopole solutions and reciprocity'', \anp {154}{1984}{253}}

\lref\opennc{N.~Nekrasov, hep-th/0203109\semi K.-Y.Kim, B.-H. Lee,
H.S. Yang, hep-th/0205010}

\lref\donaldson{S.K.~Donaldson, ``Instantons and Geometric
Invariant Theory", \cmp{93}{1984}{453-460}}

\lref\nakajima{H.~Nakajima, ``Lectures on Hilbert Schemes of
Points on Surfaces''\semi AMS University Lecture Series, 1999,
ISBN 0-8218-1956-9}
%{\tt http://www.kusm.kyoto-u.ac.jp/~nakajima/TeX.html}}

\lref\neksch{N.~Nekrasov, A.~S.~Schwarz, hep-th/9802068,
\cmp{198}{1998}{689}}

\lref\freck{A.~Losev, N.~Nekrasov, S.~Shatashvili, hep-th/9908204,
hep-th/9911099}

\lref\rkh{N.J.~Hitchin, A.~Karlhede, U.~Lindstrom, and M.~Rocek,
\cmp{108}{1987}{535}}

\lref\branek{H.~Braden, N.~Nekrasov, hep-th/9912019\semi
K.~Furuuchi, hep-th/9912047}

\lref\wilson{G.~ Wilson, ``Collisions of Calogero-Moser particles
and adelic Grassmannian", Invent. Math. 133 (1998) 1-41.}

\lref\abs{O.~Aharony, M.~Berkooz, N.~Seiberg, hep-th/9712117,
\atmp{2}{1998}{119-153}}

\lref\avatars{A.~Losev, G.~Moore, N.~Nekrasov, S.~Shatashvili,
hep-th/9509151}

\lref\abkss{O.~Aharony, M.~Berkooz, S.~Kachru, N.~Seiberg,
E.~Silverstein, hep-th/9707079, \atmp{1}{1998}{148-157}}

\lref\witsei{N.~Seiberg, E.~Witten, hep-th/9908142,
\jhep{9909}{1999}{032}}

\lref\kkn{V.~Kazakov, I.~Kostov, N.~Nekrasov, ``D-particles,
Matrix Integrals and KP hierarchy'', \np{557}{1999}{413-442},
hep-th/9810035}

\lref\DHf{J.~J.~Duistermaat, G.J.~Heckman, Invent. Math. {\bf 69}
(1982) 259\semi M.~Atiyah, R.~Bott, ``The moment map and
equivariant cohomology'', Topology {\bf 23} No 1 (1984) 1}
\lref\tdgt{M.~Atiyah, R.~Bott, ``The Yang-Mills equations over
Riemann surfaces'', Phil. Trans. Roy. Soc. London {\bf A 308}
(1982), 524--615\semi
 E.~Witten, hep-th/9204083\semi S.~Cordes,
G.~Moore, S.~Rangoolam, hep-th/9411210}

\lref\atiyahsegal{M.~Atiyah, G.~Segal, Ann. of Math. {\bf 87}
(1968) 531}

\lref\bott{R.~Bott, J.~Diff.~Geom. {\bf 4} (1967) 311}

\lref\torusaction{G.~Ellingsrud, S.A.Stromme, Invent. Math. {\bf
87} (1987) 343--352\semi L.~G\"ottche, Math. A.. {\bf 286} (1990)
193--207}

\lref\gravilit{M.~Bershadsky, S.~Cecotti, H.~Ooguri, C.~Vafa,
\cmp{165}{1994}{311}, \np{405}{1993}{279}\semi I.~Antoniadis,
E.~Gava, K.S.~Narain, T.~R.~Taylor, \np{413}{1994}{162},
\np{455}{1995}{109}}

\lref\calculus{N.~Dorey, T.~J.~Hollowood, V.~V.~Khoze,
M.~P.~Mattis, hep-th/0206063}

\lref\instmeasures{N.~Dorey, V.V.~Khoze, M.P.~Mattis,
hep-th/9706007, hep-th/9708036}

\lref\twoinst{N.~Dorey, V.V.~Khoze, M.P.~Mattis, hep-th/9607066}

\lref\vafaengine{S.~Katz, A.~Klemm, C.~Vafa, hep-th/9609239}

\lref\connes{A.~Connes, ``Noncommutative geometry'', Academic
Press (1994)}

\lref\macdonald{I.~Macdonald, ``Symmetric functions and Hall
polynomials'', Clarendon Press, Oxford, 1979}

\lref\nikfive{N.~Nekrasov, hep-th/9609219 \semi A.~Lawrence,
N.~Nekrasov, hep-th/9706025}

\lref\instlit{Literature on instantons}

\lref\givental{A.~Givental, alg-geom/9603021}

\lref\maxim{M.~Konstevich, hep-th/9405035}

\lref\whitham{A.~Gorsky, A.~Marshakov, A.~Mironov, A.~Marshakov,
hep-th/9802007}

\lref\kricheverwhitham{I.~Krichever, hep-th/9205110,
\cmp{143}{1992}{415}}

\lref\sw{N.~Seiberg, E.~Witten, hep-th/9407087, hep-th/9408099}

\lref\swsol{A.~Klemm, W.~Lerche, S.~Theisen, S.~Yankielowisz,
hep-th/9411048 \semi P.~Argyres, A.~Faraggi, hep-th/9411057\semi
A.~Hannany, Y.~Oz, hep-th/9505074}

\lref\hollowood{T.~Hollowood, hep-th/0201075, hep-th/0202197}

\lref\nsvz{V.~Novikov, M.~Shifman, A.~Vainshtein, V.~Zakharov,
\pl{217}{1989}{103}}

\lref\seibergone{N.~Seiberg, \pl{206}{1988}{75}}

\lref\dbound{G.~Moore, N.~Nekrasov, S.~Shatashvili,
hep-th/9803265}

\lref\ihiggs{G.~Moore, N.~Nekrasov, S.~Shatashvili, hep-th/9712241
}

\lref\potsdam{W.~Krauth, H.~Nicolai, M.~Staudacher,
hep-th/9803117}

\lref\kirwan{F.~Kirwan, ``Cohomology of quotients in symplectic
and algebraic geometry'', Mathematical Notes, Princeton Univ.
Press, 1985}

\lref\wittfivebrane{E.~Witten, hep-th/9610234}

\lref\issues{A.~Losev, N.~Nekrasov, S.~Shatashvili,
hep-th/9711108, hep-th/9801061}

\lref\adhm{M.~Atiyah, V.~Drinfeld, N.~Hitchin, Yu.~Manin, Phys.
Lett. {\bf 65A} (1978) 185}

\lref\warner{A.~Klemm, W.~Lerche, P.~Mayr, C.~Vafa, N.~Warner,
hep-th/9604034}

\lref\wittensolution{E.~Witten, hep-th/9703166}

\lref\twists{E.~Witten, hep-th/9304026 \semi O.~Ganor,
hep-th/9903110 \semi H.~Braden, A.~Marshakov, A.~Mironov,
A.~Morozov, hep-th/9812078}

\lref\experiment{G.~Chan, E.~D'Hoker, hep-th/9906193 \semi
E.~D'Hoker, I.~Krichever, D.~Phong, hep-th/9609041\semi
J.~Edelstein, M.~Gomez-Reino, J.~Mas, hep-th/9904087 \semi
J.~Edelstein, M.~Mari\~no, J.~Mas hep-th/9805172}

\lref\promise{N.~A.~Nekrasov and friends, to appear}

\lref\todalit{K.~Ueno, K.~Takasaki, Adv. Studies in Pure Math.
{\bf 4} (1984) 1 \semi For an excellent review see, e.g.
S.~Kharchev, hep-th/9810091}

%czj
\input Tex-document.sty

\def\firstpage{477}

\pageno=477

\def\shorttitle{Seiberg-Witten Prepotential from Instanton Counting}
\def\shortauthor{Nikita A. Nekrasov}

% \Title{\vbox{\baselineskip 10pt \hbox{}
%\hbox{ITEP-TH-22/02} \hbox{IHES/P/04/22} \hbox{hep-th/0306211}  }} {\vbox{\vskip -30 true pt
%\smallskip
%   \centerline{SEIBERG-WITTEN PREPOTENTIAL}   \smallskip\smallskip\centerline{ FROM INSTANTON COUNTING}
%\vskip4pt }} \vskip -20 true pt \centerline{Nikita A.~Nekrasov\foot{on leave of absence from ITEP, 117259, Moscow,
%Russia}}
%\smallskip\smallskip
%\smallskip\bigskip
% \centerline{\it
%Institut des Hautes Etudes Scientifiques, Le Bois-Marie, Bures-sur-Yvette, F-91440 France}
%\medskip \centerline{\tt e-mail:
%nikita@ihes.fr}
%\bigskip

%\bigskip
%\centerline{\sl To Arkady Vainshtein on his 60th anniversary}

\title{\centerline{Seiberg-Witten Prepotential}
\centerline{from Instanton Counting}}

\author{Nikita A. Nekrasov\footnote{\eightrm *}{\eightrm \hskip -5mm Institut des Hautes Etudes Scientifiques, Le Bois-Marie,
Bures-sur-Yvette, F-91440, France. On leave of absence from ITEP, 117259, Moscow, Russia. E-mail: nikita@ihes.fr}}

\centerline{\it $($To Arkady Vainshtein on his 60th anniversary$)$}

\vskip 7mm

\centerline{\boldnormal Abstract}

\vskip 4.5mm

{\narrower \ninepoint \smallskip In my lecture I consider integrals over moduli spaces of supersymmetric gauge
field configurations (instantons, Higgs bundles, torsion free sheaves ).

The applications are twofold: physical and mathematical; they involve supersymmetric quantum mechanics of
D-particles in various dimensions, direct computation of the celebrated Seiberg-Witten prepotential, sum rules for
the solutions of the Bethe ansatz equations and their relation to the Laumon's nilpotent cone. As a by-product we
derive some combinatoric identities involving the sums over Young tableaux.

}
%\vskip 4.5mm

%\noindent {\bf 2000 Mathematics Subject Classification:} 20C30, 20J05.

%{\bf Keywords and Phrases:} Cohomology, Symmetric group, Rotation group.}

\vskip 10mm

\newsec{Introduction}

The dynamics of gauge theories is a long and fascinating subject. The dynamics of supersymmetric gauge theories is
a subject which is better understood \nsvz\ yet may teach us something about the real QCD. The solution of Seiberg
and Witten \sw\ of ${\CN}=2$ gauge theory using the constraints of special geometry of the moduli space of vacua
led to numerous achievements in understanding of the strong coupling dynamics of gauge theory and also in string
theory, of which the gauge theories in question arise as low energy limits. The low energy effective Wilsonian
action for the massless vector multiplets $a$ is governed by the prepotential $\CF$, which receives one-loop
perturbative and instanton non-perturbative corrections: \eqn\prepo{{\CF} (a) = {\CF}^{pert} ( a ; {\Lambda}) +
{\CF}^{inst} (a; {\Lambda}).} In spite of the fact that these instanton corrections were calculated in many
indirect ways, their gauge theory calculation is lacking beyond two instantons\seibergone\twoinst. The problem is
that the instanton measure seems to get very complicated with the growth of the instanton charge, and the
integrals are hard to evaluate.

The present paper attempts the solution of this problem via the
localization technique, proposed long time ago in
\issues\dbound\ihiggs. Our method can be explained rather simply
in the physical terms. We calculate the vacuum expectation value
of certain gauge theory observables. These observables are
annihilated by a combination of the supercharges, and their
expectation value is not sensitive to various parameters. In
particular, one can do the calculation in the ultraviolet, where
the theory is weakly coupled and the instantons dominate. Or, one
can do the calculation in the infrared, where it is rather simple
to relate the answer to the prepotential of the effective
low-energy theory. By equating these two calculations we obtain
the desired formula.

{\bf Remark.} We can also formulate our results in a more mathematical language. We study $G \times {\bf T}^2$
equivariant cohomology of the (suitably partially compactified) moduli space $\widetilde{\CM}_{k}$ of framed
$G$-instantons on ${\IR}^4$, where $G$ is the gauge group, which acts by rotating the gauge orientation of the
instantons at infinity, and ${\bf T}^2$ is the maximal torus of $SO(4)$ -- the group of rotations of ${\IR}^4$
which also acts naturally on the moduli space. We consider the following quantity: \eqn\local{Z  (a , {\e}_1,
{\e}_2 ; q )= \sum_{k=0}^{\infty} q^k \oint_{\widetilde{\CM}_{k}} 1 } where $\oint 1$ denotes the {\it
localization} of $1$ in $H^{*}_{G \times {\bf T}^2} ( {\widetilde{{\CM}_{k}}} )$. The latter takes values in the
field of fractions of the ring $H^{*}_{G \times {\bf T}^2}(pt)$ which is identified with the space of $G \times
{\bf T}^2$ invariant polynomial functions on the Lie algebra of $G \times {\bf T}^2$. By the Chevalley theorem the
latter is isomorphic to the ring of Weyl invariant functions on the Cartan subalgebra of $G$ and ${\bf T}^2$. We
denote the coordinates on the Cartan of $G$ by $a$ and the coordinates on the Lie algebra of ${\bf T}^2$ by
${\e}_1, {\e}_2$. In explicit calculations we represent $1$ by a cohomologically equal form which allows to
replace $\oint 1$ by an ordinary integral: \eqn\repr{\oint_{\widetilde{{\CM}_k}} 1 = \int_{\widetilde{{\CM}_{k}}}
{\exp} \ {\o} + {\m}_{G} (a) + {\m}_{{\bf T}^2} ({\e}_1, {\e}_2) } where ${\o}$ is a symplectic form on
$\widetilde{{\CM}_{k}}$, invariant under the $G \times {\bf T}^2$ action, and ${\m}_{G}, {\m}_{{\bf T}^2}$ are the
corresponding moment maps.

Our first claim is \eqn\prep{\mathboxit{Z ( a, {\e}_1, {\e}_2 ; q) = {\exp} \left( {{\CF}^{inst} (a, {\e}_1,
{\e}_2; q) \over {\e}_1 {\e}_2} \right)}} where the function ${\CF}^{inst}$ is analytic in ${\e}_1, {\e}_2$ near
${\e}_1 = {\e}_2 = 0$.

We also have the following explicit expression for $Z$ in the case ${\e}_1 = - {\e}_2 = {\hbar}$ (in the general
case we also have a formula, but it looks less transparent) for $G = SU(N)$ (a simple generalization to $SO$ and
$Sp$ cases will be presented in \promise) : \eqn\explct{\mathboxit{Z ( a, {\hbar}, -{\hbar}; q) = \sum_{{\vec {\bf
k}}} q^{\vert {\bf k} \vert} \prod_{(l,i) \neq (m,j) } {{a_{lm} + {\hbar} \left( k_{l,i} - k_{m,j} +j - i
\right)}\over{a_{lm} + {\hbar} \left( j - i \right)}}.}}Here the sum is over all colored partitions: $\vec{\bf k}
= \left( {\bf k}_1, \ldots, {\bf k}_N \right)$, ${\bf k}_{l} = \{ k_{l,1} \geq k_{l,2} \geq \ldots k_{l,n_l} \geq
k_{l,n_{l}+1} =  k_{l, n_{l}+2} = \ldots = 0 \}$, and $$\vert {\vec {\bf k}} \vert = \sum_{l, i} k_{l,i}  \ ,$$
and the product is over $1\leq l,m \leq N$, and $ i,j \geq 1$.

Already \explct\ can be used to make rather powerful checks of the Seiberg-Witten solution. But the checks are
more impressive when one considers the theory with fundamental matter.

To get there one studies the bundle $V$ over ${\widetilde{{\CM}_k}}$ of the solutions of the Dirac equation in the
instanton background. Let us consider the theory with $N_f$ flavors. It can be shown that the gauge theory
instanton measure calculates in this case (cf. \instmeasures): \eqn\instmat{Z (a, m, {\e}_1, {\e}_2; q) = \sum_{k}
q^{k} \oint_{\widetilde{{\CM}_k}} {\rm Eu}_{G \times {\bf T}^2\times U(N_f)} ( V \otimes M )} where $M =
{\IC}^{N_f}$ is the flavor space, where the flavor group $U(N_f)$ acts, $m = (m_1, \ldots, m_{N_f})$ are the
masses = the coordinates on the Cartan subalgebra of the flavor group Lie algebra, and finally ${\rm Eu}_{G \times
{\bf T}^2 \times U(N_f)}$ denotes the equivariant Euler class.

The formula \explct\ generalizes in this case to: \eqn\explctm{\eqalign{Z (a, m, {\e}_1, {\e}_2; q) & = \sum_{\vec
{\bf k}} \left( q {\hbar}^{N_f} \right)^{\vert {\bf k} \vert} \prod_{(l,i)} \prod_{f=1}^{N_f} {{{\Gamma} ( {a_l +
m_f \over {\hbar}} + 1 + k_{l,i} - i )}\over{{\Gamma} ( {a_l + m_f \over \hbar} + 1 - i )}} \cr & \quad \times \
\prod_{(l,i) \neq (m,j) } {{a_{lm} + {\hbar} \left( k_{l,i} - k_{m,j} +j - i \right)}\over{a_{lm} + {\hbar} \left(
j - i \right)}}.\cr}} Again, we claim that \eqn\analytic{{\CF}^{inst}( a, m, {\e}_1, {\e}_2; q) = {\e}_1 {\e}_2 \
{\rm log} Z ( a, m, {\e}_1, {\e}_2; q)} is analytic in ${\e}_{1,2}$.

The formulae \explct\explctm\ were checked against the
Seiberg-Witten solution \swsol. Namely, we claim that
${\CF}^{inst} ( a, m , {\e}_1, {\e}_2) \vert_{{\e}_1 = {\e}_2 = 0}
= $ the instanton part of the prepotential of the low-energy
effective theory of the ${\CN}=2$ gauge theory with the gauge
group $G$ and $N_f$ fundamental matter hypermultiplets. We have
checked this claim by an explicit calculation for up to five
instantons, against the formulae in \experiment. There is also a
generalization of \explct\ to the case of adjoint matter. We shall
present it in the main body of the paper.

\newsec{Field theory expectations}

In this section we explain our approach in the field theory
language. We exploit the fact that the supersymmetric gauge theory
on flat space has a large collection of observables whose
correlation functions are saturated by instanton contribution in
the limit of weak coupling. In addition, in the presence of the
adjoint scalar vev these instantons tend to shrink to zero size.
Moreover, the observables we choose have the property that the
instantons which contribute to their expectation values are
localized in space. This solves the problem of the runaway of
point-like instantons, pointed out in \issues.

\subsec{Supersymmetries and twisted supersymmetries}

The ${\CN}=2$ theory has eight conserved supercharges, $Q_{\a}^i,
Q_{\dot a}^i$, which transform under the global symmetry group
$SU(2)_L \times SU(2)_R \times SU(2)_I$ of which the first two
factors belong to the group of spatial rotations and the last one
is the $R$-symmetry group. The indices ${\a}, {\dot\a}, i$ are the
doublets of these respective $SU(2)$ factors. The basic multiplet
of the gauge theory is the vector multiplet. It is useful to work
in the notations which make only $SU(2)_L \times SU(2)_d$ part of
the global symmetry group manifest. Here $SU(2)_d$ is the diagonal
subgroup of $SU(2)_R \times SU(2)_I$. If we call this subgroup a
``Lorentz group'', then the supercharges, superspace, and the
fermionic fields of the theory will split as follows:

\ndt{\bf Fermions}: ${\psi}_{\m}, {\chi}^{+}_{\m\n}, \eta$\semi
\ndt{\bf Superspace}: ${\t}^{\m}, {\bar \t}_{\m\n}^{+}, {\bar
\t}$\semi \ndt{\bf Superfield:} $ {\Phi} = {\phi} + {\t}^{\m}
{\psi}_{\m} + {\half} {\t}^{\m} {\t}^{\n} F_{\m\n} + \ldots $\semi
\ndt{\bf Supercharges}: $Q, Q_{\m\n}^{+}, G_{\m}$.

The supercharge $Q$ is a scalar with respect to the ``Lorentz
group'' and is usually considered as a BRST charge in the
topological quantum field theory version of the susy gauge theory.
It is conserved on any four-manifold.

In \wittenone\ E.~Witten has employed a self-dual two-form
supercharge $Q_{\m\n}^{+}$ which is conserved on K\"ahler
manifolds.

Our idea is to use other supercharges $G_{\m}$ as well. Their
conservation is tied up with the isometries of the four-manifold
on which one studies the gauge theory. Of course, the idea to
regularize the supersymmetric theory by subjecting it to the
twisted boundary conditions is very common both in physics
\twists, and in mathematics \torusaction\maxim\givental.

\subsec{Good observables: UV}

With respect to the standard topological supercharge $Q$ the observables one is usually interested in are the
gauge invariant polynomials ${\CO}^(0)_{P,x} = P({\phi}(x))$ in the adjoint scalar ${\phi}$, evaluated at
space-time point $x$, and its descendants: ${\CO}^{(k)}_{P, {\Sigma}} = \int_{\Sigma} P ( {\phi} + {\psi} + F )$,
where $\Sigma$ is a $k$-cycle. Unfortunately for $k > 0$ all such cycles are homologically trivial on ${\IR}^4$
and no non-trivial observables are constructed in such a way. One construct an equivalent set of observables by
integration over ${\IR}^4$ of a product of a closed $4-k$-form  ${\o} = {1\over{(4-k)!}} \ {\o}_{{\m}_1 \ldots
{\m}_{4-k}} {\t}^{{\m}_1} \ldots {\t}^{{\m}_{4-k}} $ and the $k$-form part of $P({\phi} + {\psi} +F)$:
\eqn\dualo{{\CO}_{P}^{\omega} = \int d^4 x d^4 {\t} \ {\o} (x, {\t}) \  P ( {\Phi} ).} Again, most of these
observables are $Q$-exact, as any closed $k$-form on ${\IR}^4$ is exact for $k
> 0$.

However, if we employ the rotational symmetries of ${\IR}^4$ and
work equivariantly, we find new observables.

Namely, consider the fermionic charge \eqn\rotch{{\tilde Q} = Q +
 E_{a} {\Omega}^{a}_{\m\n} x^{\n} G_{\m}.} Here ${\Omega}^a =
{\Omega}^{a}_{\m\n} x^{\n} {\p}_{\m}$ for $a=1 \ldots 6$ are the
vector fields generating $SO(4)$ rotations, and $E \in Lie
(SO(4))$ is a formal parameter.

With respect to the charge ${\tilde Q}$ the observables ${\CO}^{(k)}_{P,\Sigma}$ are no longer invariant (except
for ${\CO}^{(0)}_{P, 0}$ where $0 \in {\IR}^4$ is the origin, left fixed by the rotations.

However, the observables \dualo\ can be generalized to the new setup, producing a priori nontrivial ${\tilde
Q}$-cohomology classes. Namely, let us take any $SO(4)$-equivariant form on ${\IR}^4$. That is, take an
inhomogeneous differential form ${\Omega} (E)$ on ${\IR}^4$ which depends also on an auxiliary variable $E \in Lie
(SO(4))$ which has the property that for any $g \in SO(4)$: \eqn\equiv{g^* {\Omega} (E) = {\Omega} ( g^{-1} E g
)}where we take pullback defined with the help of the action of $SO(4)$ on ${\IR}^4$ by rotations. Such
$E$-dependent forms are called equivariant forms. On the space of equivariant forms acts the so-called equivariant
differential, \eqn\equivd{D = d + \iota_{V(E)}} where $V(E)$ is the vector field on ${\IR}^4$ representing the
infinitesimal rotation generated by $E$. For equivariantly closed (i.e. $D$-closed) form ${\Omega}(E)$ the
observable: \eqn\equivob{{\CO}_{P}^{{\Omega}(E)} = \int_{{\IR}^4} {\Omega} (E) \wedge P ( {\Phi}) } is ${\tilde
Q}$-closed.

Any $SO(4)$ invariant polynomial in $E$ is of course an example of the $D$-closed equivariant form. Such a
polynomial is characterized by its restriction onto the Cartan subalgebra of $SO(4)$, where it must be
Weyl-invariant. The Cartan subalgebra of $SO(4)$ is two-dimensional. Let us denote the basis in this subalgebra
corresponding to the decomposition ${\IR}^4 = {\IR}^2 \oplus {\IR}^2$ into a orthogonal direct sum of two
dimensional planes, by $({\e}_1, {\e}_2)$. Under the identification $Lie (SO(4)) \approx Lie (SU(2)) \oplus Lie
(SU(2))$ these map to $({\e}_1 + {\e}_2, {\e}_1 - {\e}_2)$.

Let us fix in addition a translationally invariant symplectic form ${\o}$ on ${\IR}^4$. Its choice breaks $SO(4)$
down to $U(2)$ -- the holonomy group of a K\"ahler manifold. Let us fix this $U(2)$ subgroup. Then we have a
moment map: \eqn\mmnt{{\m}: {\IR}^4 \longrightarrow Lie(U(2))^*, \qquad d {\m}(E) = \iota_{V(E)} {\o}, \quad E \in
Lie(U(2)).} And therefore, the form ${\o} - {\m} (E)$ is $D$-closed. One can find such euclidean coordinates
$x^{\n}$, ${\n} = 1, 2,3,4$ that the form ${\o}$ reads as follows: \eqn\sympl{{\o} = dx^1 \wedge dx^2 + dx^3
\wedge dx^4.} The Lie algebra of $U(2)$ splits as a direct sum of one-dimensional abelian Lie algebra of $U(1)$
and the Lie algebra of $SU(2)$. Accordingly, the moment map ${\m}$ splits as $( h, {\m}^1, {\m}^2, {\m}^3)$. In
the $x^\m$ coordinates \eqn\mmntc{h = \sum_{\m} \left( x^{\m} \right)^2 , \qquad {\m}^a = {\half}
{\eta}^{a}_{\m\n} x^{\m} x^{\n} ,} where ${\eta}^a_{\m\n}$ is the anti-self-dual 't Hooft symbol.

Finally, the choice of ${\o}$ also defines a complex structure on ${\IR}^4$, thus identifying it with ${\IC}^2$
with complex coordinates $z_1, z_2$ given by: $z_1 = x^1 + i x^2$, $z_2 = x^3 + i x^4$. For $E$ in the Cartan
subalgebra $H = {\m}(E)$ is given by the simple formula: \eqn\mcart{H = {\e}_1 \vert z_1 \vert^2 + {\e}_2 \vert
z_2 \vert^2.} After all these preparations we can formulate the correlation function of our interest:
\eqn\mastercorr{Z ( a, {\e}) = \langle {\exp} {1\over{(2\pi i)^2}} \int_{{\IR}^4} \left( {\o} \wedge {\Tr} \left(
{\phi} F + {\half} {\psi} {\psi} \right) - H \ {\Tr} \left( F \wedge F \right) \right)\rangle_{a} } where we have
indicated that the vacuum expectation value is calculated in the vacuum with the expectation value of the scalar
${\phi}$ in the vector multiplet given by $a \in {\bf t}$. More precisely, $a$ will be the central charge of
${\CN}=2$ algebra corresponding to the $W$-boson states.

\ndt{\bf Remarks.} 1) Note that the observable in \mastercorr\ makes the widely separated instantons suppressed.
More precisely, if the instantons form clusters around points ${\vec r}_1, \ldots, {\vec r}_l$ then they
contribute $\sim {\exp} - \sum_{m} H ({\vec r}_m )$ to the correlation function.

\noindent 2) One can expand \mastercorr\ as a sum over different instanton sectors: $$ Z ( a, {\e} ) =
\sum_{k=0}^{\infty} q^{k} Z_{k} (a, {\e})$$ where $q \sim {\Lambda}^{2N}$ is the dynamically generated scale --
for us -- simply the generating parameter.

3) The supersymmetry guarantees that \mastercorr\ is saturated by instantons. Moreover, the superspace of
instanton zero modes is acted on by a finite dimensional version of the supercharge ${\tilde Q}$ which becomes an
equivariant differential on the moduli space of framed instantons. Localization with respect to this supercharge
reduces the computation to the counting of the isolated fixed points and the weights of the action of the symmetry
groups (a copy of gauge group and $U(2)$ of rotations) on the tangent spaces. This localization can be understood
as a particular case of the Duistermaat-Heckman formula \DHf, as \mastercorr\ calculates essentially the integral
of the exponent of the Hamiltonian of a torus action (Cartan of $G$ times ${\bf T}^2$) against the symplectic
measure.

The counting of fixed points can be nicely summarized by a contour
integral (see below). This contour integral also can be obtained
by transforming the integral over the ADHM moduli space of the
observable \mastercorr\ evaluated on the instanton configuration,
by adding ${\tilde Q}$-exact terms, as in \ihiggs\dbound. It also
can be derived from Bott's formula \bott.

\subsec{Good observables: IR}

The nice feature of the correlator \mastercorr\ is it simple
relation to the prepotential of the low-energy effective theory.
In order to derive it let us think of the observable \mastercorr\
as of a slow varying changing of the microscopic coupling
constant. If we could completely neglect the fact that $H$ is not
constant, then its adding would simply renormalize the effective
low-energy scale ${\Lambda} \to {\Lambda} e^{-H}$.

However, we should remember that $H$ is not constant, and regard this renormalization as valid up to terms in the
effective action containing derivatives of $H$. Moreover, $H$ is really a bosonic part of the function ${\CH} (x,
{\t})$ on the (chiral part of) superspace (in \issues\ we considered such superspace-dependent deformations of the
theory on curved four-manifolds): $$ {\CH} ( x, {\t}) = H (x) + {\half} {\o}_{\m\n} {\t}^{\m} {\t}^{\n}. $$
Together these terms add up to the making the standard Seiberg-Witten effective action determined by the
prepotential ${\CF} (a; {\Lambda})$ to the one with the superspace-dependent prepotential \eqn\swpre{{\CF} (a;
{\Lambda} e^{-{\CH}(x, {\t})}) = {\CF} ( a ; {\Lambda} e^{-H} ) + {\o} \ {\Lambda}{\p}_{\Lambda} {\CF} (a;
{\Lambda} e^{-H}) + {\half} {\o}^2 \ \left( {\Lambda} {\p}_{ \Lambda} \right)^2 {\CF} (a; {\Lambda} e^{-H}).} This
prepotential is then integrated over the superspace (together with the conjugate terms) to produce the effective
action.

Now, let us go to the extreme infrared, that is let us scale the
metric on ${\IR}^4$ by a very large factor $t$ (keeping ${\o}$
intact). On flat ${\IR}^4$ the only term which may contribute to
the correlation function in question in the limit $t \to \infty$
is the last term in \mastercorr\ as the rest will (after
integration over superspace) necessarily contain couplings to the
gauge fields which will require some loop diagrams to get
non-trivial contractions, which all will be suppressed by inverse
powers of $t$. The last term, on the other hand, gives:

\eqn\masterir{Z(a; {\e}) = {\exp} - {1\over 8{\pi}^2}\int_{{\IR}^4} {\o} \wedge {\o} {{{\p}^2 \ {\CF} (a;
{\Lambda} e^{-H})}\over{( {\p} {\rm log} {\Lambda})^2}}  + O ({\e})} where we used the fact that the derivatives
of $H$ are proportional to ${\e}_{1,2}$. Recalling \sympl\mmntc\ the integral in \masterir\ reduces to:
\eqn\mastercorrir{Z ( a; {\e}_1, {\e}_2 ) = {\exp}\ {{\CF}^{inst}(a ; {\Lambda}) + O ({\e}) \over {{\e}_1
{\e}_2}}} where $${\CF}^{inst} (a; {\Lambda})  = \int_{0}^{\infty} {\p}_{H}^2 {\CF} (a; {\Lambda} e^{-H}) \ H {\rm
d} H, $$ thereby explaining our claim about the analytic properties of $Z$ and ${\CF}$.

\newsec{Instanton measure and its localization}

\subsec{ADHM data}

The moduli space ${\CM}_{k,N}$ of instantons with fixed framing at
infinity has dimension $4kN$. It has the following convenient
description. Take two complex vector spaces $V$ and $W$ of the
complex dimensions $k$ and $N$ respectively. These spaces should
be viewed as Chan-Paton spaces for $D(p-4)$ and $Dp$ branes in the
brane realization of the gauge theory with instantons.

Let us also denote by $L$ the two dimensional complex vector
space, which we shall identify with the Euclidean space ${\IR}^4
\approx {\IC}^2$ where our gauge theory lives.

Then the ADHM \adhm\ data consists of the following maps between the vector spaces: \eqn\sqnc{V
\longrightarrow^{\kern -.15in {\tau}} \quad V {\otimes} L \oplus W \longrightarrow^{\kern -.15in {\s}} \quad V
\otimes {\Lambda}^2 L} where \eqn\tausigma{\eqalign{ & {\tau} = \pmatrix{ B_2 \cr - B_1 \cr J}, \qquad {\s} =
\pmatrix{ B_1 & B_2 & I}, \cr & \cr & B_{1,2} \in {\rm End} (V), \ I \in {\rm Hom} ( W, V ), \ J \in {\rm Hom} (V,
W).}}

The ADHM construction represents the moduli space of $U(N)$
instantons on ${\IR}^4$ of charge $k$ as a hyperk\"ahler quotient
\rkh\  of the space of operators $(B_1, B_2, I, J)$ by the action
of the group $U(k)$ for which $V$ is a fundamental representation,
$B_{1},B_{2}$ transform in the adjoint, $I$ in the fundamental,
and $J$ in the anti-fundamental representations.

More precisely, the moduli space of proper instantons is obtained
by taking the quadruples $(B_{1,2}, I, J)$ obeying the so-called
ADHM equations: \eqn\mmntmps{{\m}_c = 0, \qquad {\m}_r = 0
,}where: \eqn\adhmeq{\eqalign{& {\m}_c = [B_1, B_2] + IJ, \cr
{\m}_r = & [B_1, B_1^{\dagger}] + [ B_2, B_2^{\dagger}] +
II^{\dagger} - J^{\dagger} J \cr}} and with the additional
requirement that the stabilizer of the quadruple in $U(k)$ is
trivial. This produces a non-compact hyperk\"ahler manifold
$M_{k,N}$ of instantons with fixed framing at infinity.

The framing is really the choice of the basis in $W$. The group $U(W) = U(N)$ acts on these choices, and acts on
$M_{k,N}$, by transforming $I$ and $J$ in the anti-fundamental and the fundamental representations respectively.

This action also preserves the hyperk\"ahler structure of $M_{k,N}$ and is generated by the hyperk\"ahler moment
maps: \eqn\glbmmps{{\bf m}_r = I^{\dagger} I - J J^{\dagger}, \qquad {\bf m}_c = J I.} Actually, ${\Tr}_{W} {\bf
m}_{r,c} = {\Tr}_{V} {\m}_{r,c}$, thus the central $U(1)$ subgroup of $U(N)$ acts trivially on $M_{k,N}$.
Therefore it is the group $G = SU(N)/{\IZ}_N$ which acts non-trivially on the moduli space of instantons.

\subsec{Instanton measure}

The supersymmetric gauge theory measure can be regarded as an infinite-dimensional version of the equivariant
Matthai-Quillen representative of the Thom class of the bundle ${\Gamma} \left( {\Omega}^{2,+} \otimes {\bf
g}_P\right)$ over the infinite-dimensional space of all gauge fields ${\CA}_{P}$ in the principal $G$-bundle $P$
(summed over the topological types of $P$). In physical terms, in the weak coupling limit we are calculating the
supersymmetric delta-function supported on the instanton gauge field configurations. In the background of the
adjoint Higgs vev, this supersymmetric delta-function is actually an equivariant differential form on the moduli
space $M_{k,N}$ of instantons. It can be also represented using the finite-dimensional hyperk\"ahler quotient ADHM
construction of $M_{k,N}$ (as opposed to the infinite-dimensional quotient of the space of all gauge fields by the
action of the group of gauge transformations, trivial at infinity) \ihiggs: \eqn\insm{ \int {\CD} {\phi} {\CD}
{\bar\phi} {\CD} {\vec H} {\CD} {\vec\chi} {\CD} {\eta} {\CD} {\Psi} {\CD} B {\CD} I {\CD} J \ e^{{\tilde Q}
\left( {\vec\chi} \cdot {\vec \m} (B, I, J) + {\Psi} \cdot V \left( {\bar\phi} \right) + {\eta} [ {\phi},
{\bar\phi} ] \right)}}where, say: \eqn\qtild{\eqalign{& {\tilde Q} B_{1,2} = {\Psi}_{B_{1,2}}, \quad {\tilde Q}
{\Psi}_{B_{1,2}} =  [ {\phi} , B_{1,2}] + {\e}_{1,2} B_{1,2}, \cr & {\tilde Q} I = {\Psi}_{I}, \quad {\tilde Q}
{\Psi}_{I} = {\phi} I - I a, \cr & {\tilde Q} J = {\Psi}_{J}, \quad {\tilde Q} {\Psi}_{J} = - J {\phi} + J a -
({\e}_1 + {\e}_2) J, \cr & {\tilde Q} {\chi}_r = H_{r}, \ {\tilde Q} H_r = [ {\phi}, {\chi}_r ], \qquad {\tilde Q}
{\chi}_c = H_c, \ {\tilde Q} H_c = [ {\phi}, \chi_{c} ] + ( {\e}_1 + {\e}_2 ) {\chi}_c, \cr & {\Psi} \cdot V
\left( {\bar\phi} \right) = {\Tr} \left( {\Psi}_{B_{1}} [ {\bar\phi}, B_{1}^{\dagger}] + {\Psi}_{B_{2}} [
{\bar\phi}, B_{2}^{\dagger}] + {\Psi}_{I} [{\bar\phi}, I^{\dagger}] - {\Psi}_{J} [ {\bar \phi}, J^{\dagger}] +
c.c. \right) \cr}} (we refer to \ihiggs\ for more detailed explanations). If the moduli space $M_{k,N}$ was
compact and smooth one could interpret \insm\ as a certain topological quantity and apply the powerful equivariant
localization techniques \tdgt\ to calculate it.

The non-compactness of the moduli space of instantons is of both ultraviolet and of infrared nature. The UV
non-compactness has to do with the instanton size, which can be made arbitrarily small. The IR non-compactness has
to do with the non-compactness of ${\IR}^4$ which permits the instantons to run away to infinity.

\subsec{Curing non-compactness}

The UV problem can be solved by relaxing the condition on the stabilizer, thus adding the so-called point-like
instantons. A point of the hyperk\"ahler space ${\tilde M}_{k,N}$ with orbifold singularities which one obtains in
this way (Uhlenbeck compactification) is an instanton of charge $p \leq k$ and a set of $k-p$ points on ${\IR}^4$:
\eqn\uhl{{\tilde M}_{k,N} = M_{k,N} \cup M_{k-1,N} \times {\IR}^4 \cup M_{k-2,N} \times Sym^2 ({\IR}^4) \cup
\ldots \cup Sym^k ({\IR}^4).}

The resulting space ${\tilde M}_{k,N}$ is a geodesically complete
hyperk\"ahler orbifold. Its drawback is the non-existence of the
universal bundle with the universal instanton connection over
${\tilde M}_{k,N} \times {\IR}^4$. We actually think that in
principle one can still work with this space. Fortunately, in the
case of $SU(N)$ gauge group there is a nice space
$\widetilde{\CM}_{k,N}$ which is obtained from ${\tilde M}_{k,N}$
by a sequence of blowups (resolution  of singularities) which is
smooth, and after some modification of the gauge theory
(noncommutative\connes\cds\douglashull\witsei\ deformation)
becomes a moduli space with the universal instanton. Technically
this space is obtained \neksch\ by the same ADHM construction
except that now one performs the hyperk\"ahler quotient at the
non-zero level of the moment map:
$$
{\m}_r = {\z}_r {\bf 1}_{V}, \qquad {\m}_c = 0
$$
(one can also make ${\m}_c \neq 0$ but this does not give anything
new). The cohomology theory of $\widetilde{\CM}_{k,N}$ is richer
then that of ${\tilde M}_{k,N}$ because of the exceptional
divisors. However, our goal is to study the original gauge theory.
Therefore we are going to consider the (equivariant) cohomology
classes of $\widetilde{\CM}_{k,N}$ lifted from ${\tilde M}_{k,N}$.

As we stated in the introduction, we are going to utilize the
equivariant symplectic volumes of $\widetilde{\CM}_{k,N}$. This is
not quite precise. We are going to consider the symplectic
volumes, calculated using the closed two-form lifted from ${\tilde
M}_{k,N}$. This form vanishes when restricted onto the exceptional
variety. This property ensures that we don't pick up anything not
borne in the original gauge theory (don't pick up freckle
contribution in the terminology of \freck).

The ADHM construction from the previous section gives rise to the
instantons with fixed gauge orientation at infinity (fixed
framing). The group $G = SU(N)/{\IZ}_N$ acts on their moduli space
${\CM}_{N,k}$ by rotating the gauge orientation. Also, the group
of Euclidean rotations of ${\IR}^4$ acts on ${\CM}_{N,k}$. We are
going to apply localization techniques with respect to both of
these groups.

In fact, it is easier to localize first with respect to the groups
$U(k) \times G \times {\bf T}^2$ acting on the vector space of
ADHM matrices, and then integrate out the $U(k)$ part of the
localization multiplet, to incorporate the quotient.

The action of ${\bf T}^2$ is free at ``infinities'' of
${\widetilde {\CM}_{k}}$, thus allowing to apply localization
techniques without worrying about the IR non-compactness.
Physically, the integral \mastercorr\ is Gaussian-like and
convergent in the IR region.

\subsec{Reduction to contour integrals}

After all the manipulations as in \ihiggs\dbound\ we end up with the following integral\freck: \eqn\cntr{Z_{k}( a,
{\e}_{1}, {\e}_2)  = {1\over{k!}} {{\e}^k\over{(2{\pi} i {\e}_1 {\e}_2)^k}} \oint {\prod}_{i=1}^{k} {{\rm
d}{\phi}_i \ Q({\phi}_i) \over {P({\phi}_i) P({\phi}_i + {\e})}} \ {\prod}_{1\leq i < j \leq k} {{{\phi}_{ij}^2 (
{\phi}_{ij}^2 - {\e}^2)}\over{({\phi}_{ij}^2 - {\e}_1^2)({\phi}_{ij}^2- {\e}_2^2)}}} where: \eqn\poly{\eqalign{&
Q(x) = \prod_{j=1}^{N_f} ( x + m_j), \cr & P (x) = {\prod}_{l=1}^{N} ( x - a_l), \cr}}${\phi}_{ij}$ denotes
${\phi}_i - {\phi}_j$ and ${\e} = {\e}_1 + {\e}_2$.

We went slightly ahead of time and presented the formula which
covers the case of the gauge theory with $N_f$ fundamental
multiplets. In fact, the derivation is rather simple if one keeps
in mind the relation to the Euler class of the Dirac zeromodes
bundle over the moduli space of instantons, stated in the
introduction.

\subsec{Classification of the residues}

The integrals \cntr\ should be viewed as contour integrals. As
explained in \dbound\ the poles at ${\phi}_{ij} = {\e}_1, {\e}_2$
should be avoided by shifting ${\e}_{1,2} \to {\e}_{1,2} + i0$,
those at ${\phi}_i = a_l$ similarly by $a_l \to a_l + i0$ (this
case was not considered in \dbound\ but actually was considered
(implicitly) in \ihiggs). The interested reader should consult
\kirwan\ for more mathematically sound explanations of the contour
deformations arising in the similar context in the study of
symplectic quotients.

The poles which with non-vanishing contributions to the integral must have ${\phi}_{ij} \neq 0$, for $i \neq j$,
otherwise the numerator vanishes. This observation simplifies the classification of the poles. They are labelled
as follows. Let $k = k_1 + k_2 + \ldots + k_N$ be a partition of the instanton charge in $N$ summands which have
to be non-negative (but may vanish), $k_l \geq 0$. In turn, for all $l$ such that  $k_l > 0$ let $Y_l$ denote a
partition of $k_l$: $$ k_l = k_{l,1} + \ldots k_{l, {\n}^{l,_1}}, \qquad k_{l,1} \geq k_{l,2} \geq \ldots \geq
k_{l, {\n}^{l,1}} > 0. $$ Let ${\n}^{l,1} \geq {\n}^{l,2} \geq \ldots {\n}^{l,k_{l,1}}>0$ denote  the dual
partition. Pictorially one represents these partitions by the Young diagram with $k_{l,1}$ rows of the lengths
${\n}^{l,1}, \ldots {\n}^{l,k_{l,1}}$. This diagram has ${\n}^{l,1}$ columns of the lengths $k_{l,1}, \ldots,
k_{l,{\n}^{l,1}}$.

In total we have $k$ boxes distributed among $N$ Young tableaux
(some of which could be empty, i.e. contain zero boxes). Let us
label these boxes somehow (the ordering is not important as it is
cancelled in the end by the factor $k!$ in \cntr). Let us denote
the collection of $N$ Young diagrams by ${\vec Y} = (Y_1, \ldots,
Y_N)$. We denote by $\vert Y_l \vert = k_l$ the number of boxes in
the $l$'th diagram, and by $\vert \vec Y \vert = \sum_l \vert Y_l
\vert = k$.

Then the pole of the integral \cntr\ corresponding to $\vec Y$ is at ${\phi}_s$ with $s$ labelling the box
$({\a},{\b})$ in the $l$'th Young tableau (so that $0 \leq {\a} \leq {\n}^{l,{\b}}, \ 0 \leq {\b} \leq
k_{l,{\a}}$) equal to: \eqn\pole{{\vec Y} \longrightarrow  {\phi}_{s} = a_l + {\e}_1 ({\a}-1) + {\e}_2 ( {\b}-1).}

\subsec{Residues and fixed points}

The poles in the integral \cntr\ correspond to the fixed points of
the action of the groups $G \times {\bf T}^2$ on the resolved
moduli space ${\tilde\CM}_{k,N}$. Physically they correspond to
the $U(N)$ (noncommutative) instantons which split as a sum of
$U(1)$ noncommutative instantons corresponding to $N$ commuting
$U(1)$ subgroups of $U(N)$. The instanton charge $k_l$ is the
charge of the $U(1)$ instanton in the $l$'th subgroup. Moreover,
these abelian instantons are of special nature -- they are fixed
by the group of space rotations. If they were commutative (and
therefore point-like) they had to sit on top of each other, and
the space of such point-like configurations would have been rather
singular. Fortunately, upon the noncommutative deformation the
singularities are resolved. The instantons cannot sit quite on top
of each other. Instead, they try to get as close to each other as
the uncertainty principle lets them. The resulting abelian
configurations were classified (in the language of torsion free
sheaves) by H.~Nakajima \nakajima.

Now let us fix a configuration ${\vec Y}$ and consider the
corresponding contribution to the integral over instanton moduli.
It is given by the residue of the integral \cntr\ corresponding to
\pole:

\eqn\res{\eqalign{&R_{\vec Y} =  {1\over{({\e}_1 {\e}_2)^k}} \prod_{l} \prod_{{\a}=1}^{{\n}^{l,1}}
\prod_{{\b}=1}^{k_{l,{\a}}} {{{\CS}_{l}({\e}_1 ({\a}-1) + {\e}_2 ({\b}-1))}\over{({\e} ({\ell}(s)+1) - {\e}_2
h(s))({\e}_2 h(s) - {\e} {\ell} (s))}} \times \cr & \prod_{l < m} \prod_{{\a}=1}^{{\n}^{l,1}}
\prod_{{\b}=1}^{k_{m,1}} \left( {{\left( a_{lm} + {\e}_1 ({\a} - {\n}^{m,{\b}})+{\e}_2 (1-{\b}) \right) \left(
a_{lm} + {\e}_1 {\a} + {\e}_2 ( k_{l,\a} + 1 - {\b}) \right)}\over{\left( a_{lm} + {\e}_1 {\a} + {\e}_2 ( 1 -
{\b}) \right) \left( a_{lm} + {\e}_1 ({\a} - {\n}^{m,{\b}}) + {\e}_2 ( k_{l,{\a}} + 1 - {\b}) \right)}}\right)^2
\cr}} where we have used the following notations: $a_{lm} = a_l - a_m$,  \eqn\sss{{\CS}_{l}(x) = {Q(a_l + x) \over
{\prod_{m \neq l} ( x + a_{lm} ) (x + {\e} + a_{lm} )} }, \qquad S_{l}(x) = {Q(a_l + x)\over \prod_{m \neq l} ( x
+ a_{lm})^2} , } and \eqn\hkl{{\ell}(s) = {k}_{l,{\a}} - {\b}, \qquad h(s) = {k}_{l,{\a}} + {\n}^{l,{\b}} - {\a} -
{\b} +1.}

Now, if we set ${\e}_1 = {\hbar} = - {\e}_2$ the formula \res\ can
be further simplified. After some reshuffling of the factors it
becomes exactly the summand in \explctm.

\subsec{The first three nonabelian instantons}

We shall now give the formulae for the first three instanton
contributions to the prepotential for the general $SU(N)$ case,
with $N_f < 2N$.

We shall work with ${\e}_1 = {\hbar} = - {\e}_2$. It will be
sufficient to derive the gauge theory prepotential.

Directly applying the rules \cntr\res\ we arrive at the following expressions for the moduli integrals:
\eqn\zeds{\eqalign{Z_1 & = {1\over{{\e}_1 {\e}_2}} \sum_{l} S_{l}, \cr Z_2 & = {1\over{( {\e}_1 {\e}_2)^2}} \left(
{1\over 4} \sum_{l} S_{l} \left( S_l ( +\hbar) + S_{l}( - \hbar) \right) + {1\over 2} \sum_{l \neq m} {S_{l} S_{m}
\over \left( 1 - {{\hbar}^2 \over a_{lm}^2} \right)^{2} }\right), \cr Z_{3} & = {1\over{({\e}_1 {\e}_2)^3}}
\Biggl( \sum_{l} {S_{l}  \left( S_{l} (+ \hbar) S_{l}(+ 2 \hbar) + S_{l}( - \hbar) S_{l} (- 2\hbar) + 4 S_{l}( +
\hbar) S_{l}(- \hbar) \right)\over 36} \cr  & \quad + \sum_{l \neq m} {S_{l} S_{m} \over 4} \left( {S_{l} (+ \hbar
) \over \left( 1 - {2 \hbar^2 \over ( a_{lm} ( a_{lm} + \hbar ))} \right)^{2} } + {S_{l}(- \hbar) \over \left( 1 -
{2 \hbar^2 \over ( a_{lm} ( a_{lm} - \hbar))} \right)^{2}} \right) \cr & \quad + \sum_{l \neq m \neq n} {S_{l}
S_{m} S_{n} \over 6 \left( \left( 1 - {\hbar^2 \over a_{lm}^2 } \right)\left( 1 - {\hbar^2 \over a_{ln}^2 }
\right)\left( 1 - {\hbar^2 \over a_{mn}^2 } \right) \right)^{2}} \Biggr), \cr}} from which we immediately derive:
\eqn\prep{\eqalign{& {\CF}_1 = \sum_{l} S_{l}, \cr & {\CF}_2 = \sum_{l} {1\over 4} S_{l} S_{l}^{(2)} + \sum_{ l
\neq m} {S_{l} S_{m}  \over a_{lm}^2}  + O({\hbar}^2), \cr & {\CF}_3 = \sum_{l} {S_{l} \over 36} \left( S_{l}
S_{l}^{(4)} + 2 S_{l}^{(1)} S_{l}^{(3)} + 3 S_{l}^{(2)} S_{l}^{(2)} \right) \cr & \qquad + \sum_{l \neq m} {S_{l}
S_{m} \over a_{lm}^4}\left( 5 S_{l}  - 2 a_{lm} S_{l}^{(1)} + a_{lm}^2 S_{l}^{(2)} \right) \cr & \qquad
+\sum_{l\neq m\neq n} {2 S_{l} \ S_{m} S_{n} \over 3 ( a_{lm} a_{ln} a_{mn})^2 } \left(a_{ln}^2 + a_{lm}^2 +
a_{mn}^2 \right)  + O ({\hbar}^2). \cr}}

\subsec{Four and five instantons}

To collect more ``experimental data-points'' we have considered
the case of the gauge groups $SU(2)$ and $SU(3)$ with fundamental
matter. We have computed explicitly the prepotential for four and
five instantons and found  a perfect agreement (yet a few typos)
with the results of \experiment. We should stress that this is a
non-trivial check. Just as an example, we quote here the
expression for ${\CF}_5$: $${\CF}_5 (a, m) = {{{\m}_3}\over{8
a^{18}}} ( 35 a^{12} - 210 a^{10} {\m}_2 + a^8 \left( 207 {\m}_2^2
+ 846 {\m}_4 \right) $$ $$  - 1210 a^6 {\m}_2 {\m}_4 + a^4 \left(
1131 {\m}_4^2 + 3698 {\m}_3^2 {\m}_2 \right) - 5250 a^2 {\m}_3^2
{\m}_4 + 4471 {\m}_3^4 ),$$ where $2a = a_1 - a_2$, ${\m}_2 =
m_1^2 + m_2^2 + m_3^2$, ${\m}_3 = m_1 m_2 m_3$, ${\m}_4 = (m_1
m_2)^2 + (m_2 m_3)^2 + (m_1 m_3)^2$.

\subsec{Adjoint matter and other matters}

So far we were discussing ${\CN}=2$ gauge theories with matter in
the fundamental representations. Now we shall pass to other
representations. It is simpler to start with the adjoint
representation. The ${\e}$-integrals \cntr\ reflect  both the
topology of the moduli space of instantons and also of the matter
bundle.

The latter is the bundle of the Dirac zero modes in the
representation of interest. For the adjoint representation, and on
${\IR}^4$ this bundle can be identified with the tangent bundle to
the moduli space of instantons. It  has a $U(1)$ symmetry. The
equivariant Euler class of the tangent bundle (= the Chern
polynomial) is the instanton measure in the case of massive
adjoint matter. This reasoning leads to the following
${\e}$-integral: \eqn\adjcntr{\eqalign{Z_{k} & = {1 \over k!}
\left( {({\e}_1 + {\e}_2)( {\e}_1 + m ) ({\e}_2 + m) \over 2{\pi}
i \ {\e}_1 {\e}_2 \ m \ ( {\e} - m)} \right)^{k} \oint
{\prod_{i=1}^{k} }{{{\rm d}{\phi}_i} \ P ({\phi}_i + m)  P (
{\phi}_i + {\e} -m) \over \ \ \ P ({\phi}_i ) P ({\phi}_i + {\e}
)} \cr & \quad \times \prod_{i < j} {{\phi}_{ij}^2 ({\phi}_{ij}^2
- {\e}^2) ({\phi}_{ij}^2 - ( {\e}_1 - m)^2)({\phi}_{ij}^2 -
({\e}_2 - m)^2) \over ({\phi}_{ij}^2 - {\e}_1^2 )({\phi}_{ij}^2 -
{\e}_2^2)({\phi}_{ij}^2 - m^2 )({\phi}_{ij}^2 - ({\e}-m)^2)}.
\cr}} Note the similarity of this expression to the contour
integrals appearing\dbound\ in the calculations of the bulk
contribution to the index of the supersymmetric quantum mechanics
with $16$ supercharges (similarly, \cntr\ is related to the one
with $8$ supercharges). This is not an accident, of course.

Proceeding analogously to the pure gauge theory case we arrive at the following expressions for the first two
instanton contributions to the prepotential (which agrees with \experiment): \eqn\adjprep{\eqalign{{\CF}_1 & = m^2
\sum_{l} T_{l}, \cr {\CF}_2 & = \sum_l \left( - {3m^2 \over 2} T_l^2 + {m^4 \over 4} T_l T_{l}^{(2)} \right) \cr &
\quad + m^4 \sum_{l \neq n} T_l T_n \left( {1\over a_{ln}^2} - {1\over 2 (a_{ln} + m)^2} - {1\over 2(a_{ln} -
m)^2} \right), \cr}} where $T_l (x) = \prod_{n \neq l} \left( 1 - {m^2 \over (x + a_{ln})^2} \right)$, $T_l =
T_l(0)$, $T_l^{(n)} = {\p}_x^n T_l(x) \vert_{x=0}$ (cf. \hollowood).

For generic representation $R$ of the gauge group we should use
the equivariant Euler class of the corresponding (virtual) vector
bundle ${\CE}_{R}$ over the moduli space of instantons \promise.

\subsec{Perturbative part}

So far we were calculating the nonperturbative part of the
prepotential. It would be nice to see the perturbative part
somewhere in our setup, so as to combine the whole expression into
something nice.

One way is to calculate carefully the equivariant Chern character
of the tangent bundle to ${\widetilde{\CM}_k}$ along the lines
sketched in the end of the previous section\promise. The faster
way in the ${\e}_1 + {\e}_2 = 0$ case is to note that the
expression \explct\ is a sum over partition with the universal
denominator, which is not well-defined without the non-universal
numerator. Nevertheless, let us try to pull it out of the sum.

We get the infinite product (up to an irrelevant constant): $$ \prod_{i,j = 1}^{\infty} \prod_{l \neq  m}
{1\over{a_{lm} + {\hbar} (i - j)}} \sim $$ \eqn\schwinger{ {\exp} - \sum_{l \neq m} \int_{0}^{\infty} {{ds} \over
s} {e^{- s a_{lm}} \over{(e^{\hbar s}- 1) ( e^{-\hbar s} -1)}}.} If we regularize this by cutting the integral at
$s \to 0$, we get a finite expression, which actually has the form $${\exp} {{\CF}^{pert} ( a , {\e}_1, {\e}_2)
\over{{\e}_1 {\e}_2}}, $$ with ${\CF}^{pert}$ being analytic in ${\e}_1, {\e}_2$ at zero. In fact $${\CF}^{pert}
(a, 0, 0) = \sum_{l \neq m} {1\over 2} a_{lm}^2 {\rm log} \ a_{lm} + {\rm  \ ambiguous \ quadratic \ polynomial \
in} \ a_{lm} . $$
 The formula \schwinger\ is a
familiar expression for the Schwinger amplitude of a mass $a_{lm}$
particle in the electromagnetic field \eqn\emf{F \propto {\e}_1 \
dx^1 \wedge dx^2 + {\e}_2 \ dx^3 \wedge dx^4 \ .} Its appearance
be explained in the next section.

Let us now combine ${\CF}^{inst}$ and ${\CF}^{pert}$ into a single ${\e}$-deformed prepotential $$ {\CF} (a,
{\e}_1, {\e}_2) = {\CF}^{pert} (a, {\e}_1, {\e}_2 ) + {\CF}^{inst} (a, {\e}_1, {\e}_2)$$ where in general we
define: \eqn\schw{{\CF}^{pert} (a, {\e}_1, {\e}_2 ) = \sum_{l \neq m} \int_{{\ve}}^{\infty} {{ds \over s}} {e^{- s
a_{lm} } \over  {\rm sinh} \left( {s {\e}_1 \over 2} \right) {\rm sinh} \left( {s {\e}_2 \over 2 } \right) }} with
the singular in ${\ve}$ part dropped. We define: \eqn\prtn{{\CZ} ( a, {\e}_1, {\e}_2 ; q) = {\exp} {{\CF} (a,
{\e}_1, {\e}_2 ; q) \over {{\e}_1 {\e}_2}}.}

\font\boldgreek = cmmib10 scaled 1440

\newsec{{\boldgreek \char28}-function conjecture}

This conjecture relates the expansion \explct\ to the dynamics of
the Seiberg-Witten curve.

Consider the theory of a free complex chiral fermion ${\psi}, {\psi}^*$,  \eqn\frferm{{\CS} = \int_{\Sigma}
{\psi}^* {\pb} {\psi}} living on the curve $\Sigma$: \eqn\swcurv{w + {{\Lambda}^{2N} \over w} ={\bf P} ({\l}),
\qquad {\bf P} ({\l}) = \prod_{l=1}^{N} ( {\l} - {\a}_l )} embedded into the space ${\IC} \times {\IC}^*$ with the
coordinates $({\l}, w)$. This curve has two distinguished points $w = 0$ and $w = \infty$ which play a prominent
role in the Toda integrable hierarchy \todalit. Let us cut out small disks $D_{0}$ and $D_{\infty}$ around these
two points.

The path integral on the surface ${\Sigma}$ with two discs deleted will give a state in the tensor product
${\CH}_0 \otimes {\CH}^{*}_{\infty}$ of the Hilbert spaces ${\CH}_0$, ${\CH}_{\infty}$ associated to ${\p}D_0$ and
${\p}D_{\infty}$ respectively. It can also be viewed as an operator $G_{\Sigma} : {\CH}_0 \to {\CH}_{\infty}$.

Choose a vacuum state $\vert 0 \rangle \in {\CH}_0$ and its dual $\langle 0 \vert \in {\CH}_{\infty}^*$ (we use
the global coordinate $w$ to identify ${\CH}_0$ and ${\CH}_{\infty}$). Consider \eqn\tu{{\tau}_{\Sigma}   =
\left\langle 0 \left\vert {\exp} \left( {1\over {\hbar}} \oint_{{\p}D_\infty} \ S \ J \right) \ G_{\Sigma} \
{\exp} \left( - {1\over{\hbar}} \oint_{{\p}D_{0}} S \ J \right) \right\vert 0 \right\rangle} where:
\eqn\not{\eqalign{&  J = : {\psi}^* {\psi} : \cr & dS = {1\over{2\pi i}} {\l} {dw \over w} \cr}}and  we choose the
branch of $S$ near $w = 0, \infty$ such that (cf. \whitham) : $$S = {N\over{2\pi i}} {w}^{\mp 1\over N} + O (
{\l}^{-1} ). $$ Let us represent $\Sigma$ as a two-fold covering of the ${\l}$-plane. It has branch points at
${\l} = {\a}^{\pm}_l$ where $${\bf P} ( {\a}^{\pm}_l ) = \pm 2 \Lambda^{N}. $$ Let us choose the cycles $A_l$ to
encircle the cuts between ${\a}_l^{-}$ and ${\a}_{l}^{+}$. Of course, in $H_1 ({\Sigma}, {\IZ})$, $\sum_l A_l =
0$. Then, we define: $$a_l =  \oint_{A_l} dS \ .$$Our final conjecture states: \eqn\fnc{\mathboxit{{\CZ} (a,
{\hbar}, - {\hbar}) = {\tau}_{\Sigma}.}} Note that from this conjecture the fact that ${\CF}_{0} (a, 0,0) $
coincides with the Seiberg-Witten expression follows as a consequence of the Krichever universal formula
\kricheverwhitham. The remaining paragraph is devoted to the explanation of the motivation behind \fnc.

Let us assume that we are in the domain where ${\a}_l - {\a}_m \gg
{\Lambda}$. Then the surface ${\Sigma}$ can be decomposed into two
halves ${\Sigma}_{\pm}$ by $N$ smooth circles $C_l$ which are the
lifts to $\Sigma$ of the cuts connecting ${\a}_l^{-}$ and
${\a}_l^{+}$. The path integral calculating the matrix element
\tu\ can be evaluated by the cutting and sewing along the $C_l$'s.
The path integral on ${\Sigma}_{\pm}$ gives a state in
$$\otimes_{l=1}^{N} {\CH}_{C_l}$$ (its dual). If we first pull the
$\oint S J$ as close to $C_l$ as possible, we shall get the Hilbert space obtained by quantization of the fermions
which have $a_l + {\half} mod {\IZ}$ moding: \eqn\psin{{\psi} (w) \sim \sum_{i \in {\IZ}} {\psi}_{l,i} w^{a_l + i}
\left( {dw \over  w} \right)^{\half}}near $C_l \subset \Sigma$. In addition, the states in ${\CH}_{C_l}$ of fixed
total $U(1)$ charge are labelled by the partitions $k_{l,i}$. We conjecture, that \eqn\fncc{\left\langle 0
\left\vert  e^{\oint S J} \prod_{l,i} {\psi}_{l,k_{l,i} - i} {\psi}^{*}_{l,-i} \right\vert 0 \right\rangle_{l}
\sim \prod_{(l,i) < (m,j)} ( a_{lm} + {\hbar} ( k_{l,i} - k_{m,j} + j - i) ).} It is clear that \fncc\ implies
\fnc. For $N=1$ \fncc\ is of course a well-known fact (with the coefficient given by $\prod_{i < j}
{1\over{j-i}}$). It gives rise to the formula (which can also be derived using the Schur identities \macdonald):
$$
Z_{N=1}({\hbar, - \hbar}) = e^{- {1\over {\hbar}^2}}
$$
which confirms that in spite of the fact that we worked with the
resolved moduli space $\cup_k \widetilde{\CM}_{k,1} =\cup_k
({\IC}^2)^[k]$ the ``symplectic'' volume we calculated is that of
$\cup_k {\tilde M}_{k,1} = \cup_k {Sym}^k ({\IR}^4)$.

\footatend%\vfill
\supereject\immediate\closeout\rfile\writestoppt
\baselineskip=14pt \noindent {{\boldLARGE References}}\bigskip{\frenchspacing%
\parindent=20pt\escapechar=` \input refs.tmp\vfill\eject}\nonfrenchspacing \bye